\documentclass[utf8]{frontiersSCNS} 

\usepackage{url,hyperref,lineno,microtype,subcaption}
\usepackage[onehalfspacing]{setspace}
\usepackage{placeins}

\usepackage{booktabs}
\usepackage{color}
\usepackage{threeparttable}

\def\keyFont{\fontsize{8}{11}\helveticabold }
\def\firstAuthorLast{Lee {et~al.}} 
\def\Authors{Hyowon Lee\,$^{1,*}$, Mingming Liu\,$^{1}$, Michael Scriney\,$^{1}$ and Alan F. Smeaton\,$^{1}$}
\def\Address{$^{1}$Insight SFI Research Centre for Data Analytics, Dublin City University, Dublin 9, Ireland }
\def\corrAuthor{Corresponding Author}

\def\corrEmail{hyowon.lee@dcu.ie}

\begin{document}
\onecolumn
\firstpage{1}

\title[Playback-centric visualisations of video usage]{Playback-centric visualisations of video usage using  weighted interactions to guide where to watch in an educational context}

\author[\firstAuthorLast ]{\Authors} 
\address{} 
\correspondance{} 

\extraAuth{}

\maketitle

\begin{abstract}



The steady increase in use of online educational tools and services has led to a large amount of educational video materials made available for students to watch. Finding the right video content is usually supported by the overarching learning management system and its user interface that organises various video items by course, categories and weeks, and makes them searchable. However, once a wanted video is found, students are often left without further guidance as to  what parts in that video they should focus on.

In this article, an additional timeline visualisation to augment the conventional playback timeline is introduced which employs a novel playback weighting strategy in which the history of different video interactions generate  scores based on the context of each playback. This includes whether the playback started after jumping forward or backward in the video, whether the playback was at a faster or slower speed, and whether the playback window was in focus on the student's screen or was in the background. The resultant scores are presented  on the additional timeline, making it in effect a playback-centric usage graph nuanced by how each playback was executed. Students are informed by this visualisation on the playback by their peers and can selectively watch those portions which the contour of the usage visualisation suggests. 

The visualisation was implemented as a fully-fledged web application and deployed in an undergraduate course at a university for two full semesters. A total of 270 students used the system throughout both semesters watching 52 videos, guided by  visualisations on what to  watch. Analysis of playback logs revealed that students selectively watched portions in videos corresponding to the most important portions of the videos as assessed by the instructor who created the videos. The characteristics of this method as a way of guiding  students as to where to watch as well as a complementary tool for playback analysis, are discussed.  Further insights into the potential values of this visualisation and its underlying playback weighting strategy are also discussed.

\tiny
 \keyFont{ \section{Keywords:} learning analytics, online learning, educational video, interaction logging, system deployment} 
\end{abstract}

\section{Introduction}

Increasing amounts of educational video materials are becoming available online. Coursera, one of the most popular Massive Open Online Course (MOOC) platforms, has over 3,000 online courses with 82 million registered learners as of March 2022 (https://about.coursera.org/), probably with millions of video materials of varying lengths available. When the COVID-19 pandemic hit in Spring 2020 forcing almost all universities to deliver their courses remotely, the amount of video content for students to remotely access further increased, with two main types of video content: (1) synchronous recorded lecture videos made available after the online, live lectures on Zoom or other video conferencing platform (typically lengthy, e.g. 30 minutes to 1 hour or longer), and (2)  video material prepared by the lecturer out of class and made available for students to watch asynchronous (typically shorter, e.g. 5 minutes to 15 minutes, but can also  be longer). 

Regardless of the type of  video material, online courses meant a considerable amount of videos for students to wade through and watch, often multiple video items per week, including lengthy lecture recordings as well as short videos to review. These videos typically become available on the university's Learning Management System (LMS) so that the videos for each week to watch are grouped, arranged, ordered and possibly searchable for students to access: students are generally provided with some ways to find a suitable subset of videos that could satisfy their learning needs.

Once a video is selected and opened, however,  students are usually left  watching the contents without much support: (i) the inherent temporal nature of the video medium hides  the full content from being browsed, and (ii) the typically unprofessionally- and minimally-edited, linearly organised characteristics of educational contents (mixtures of PowerPoint slides sometimes with animated annotations and/or talking heads, etc.) mean no clear or explicit structure to be followed within the videos. This results in either having to watch the full video even when not all of the contents are relevant/useful for the student, or blindly jumping back and forth hoping a useful point in the video will show up. Without instrumenting a lot of sophisticated within-video content browsing facilities (storyboards, keyframe-based browsing, structured shot/scene navigation, etc.) often the subject of experimentation within the video retrieval community \citep{Smeaton2004}, how could  students be guided as to where to watch inside this type of videos?

Current  analytic studies of video usage in the educational setting usually target \emph{the instructors} as the beneficiary of the analytics, i.e. to show  instructors how  students are watching the videos, as there is a clear need for more objective insights and  better visibility of students' viewing behaviour to the teachers so that they can improve how they make and provide their video materials \citep{Fong2019}.

In this paper, we shift the focus of analytics \emph{from the instructor to the students}, i.e. the usage of video contents watched by the student cohort can be analysed and the results can be used to support students' navigation within the video content, an aspect which is missing in today's educational video platforms. By focusing on how potentially useful usage information could be fed back to the students so that it will guide them in choosing where to watch within each video, a \emph{playback-centric visualisation} was devised in which we developed a playback weighting strategy that takes into account various micro-level video interactions that students trigger and that are weighted differently. These weights are used to calculate an overall score for every 1-second window within a video, and we  then present this as a video timeline.

By computing the cumulative values of individual video interactions (e.g. play/pause clicks or seek actions), and presenting the cumulative values of playback derived from differently weighted video interactions, the unique timeline visualisation represents those parts within each video that have received the  most interaction from students as it reflects how they watched when they were watching e.g. were they skipping or perhaps using faster (1.25x, 1.5x, 2x) playback, as well as when during the semester they were watching the video. Our paper introduces this playback-centric visualisation of video usage, in which:

\begin{itemize}
    \item a playback weighting strategy is used and updated daily to calculate the second-by-second scores in each video based on judiciously chosen weightings for micro-level video interactions from students;
    \item a full-fledged system incorporating this strategy was built and deployed for two full semesters (2021 and 2022) at a university, and
    \item we analyse the usage data and the usefulness of this method, and share the insights gained from it.
\end{itemize}

The contributions of the paper are that by going through the above activities, we establish our understanding of the designed visualisation, which can be expressed as our attempt at answering the following research question:

{\leftskip=2cm\relax
 \rightskip=2cm\relax
  \noindent\emph{Does the proposed visualisation of weighted playback by peers help  students to selectively watch the more important parts within  videos?}
 \par}

More analyses of the usage, perceptions and attitudes of students on an exclusively online learning environment during the pandemic \citep{Popa2020} \citep{Coman2020} \citep{Chen2020} and attempts at capturing the students' attention level and monitoring it while they watch video or live lectures \citep{Lee2021} \citep{IntelliEye2018} \citep{Srivastava2021} \citep{Pham2015} are all very important, and the support for better guidance for students in more selectively/efficiently reviewing the video contents is a notable gap identified and addressed in this paper.

\section{Related Work}
There is a large body of literature in learning analytics for online platforms in general e.g. comprehensive reviews including \citep{Ihantola2015} and \citep{Dawson2019}, and a moderate amount of literature on analysing video watching behaviour in an educational context. When it comes to detailed analysis looking into the within-video-level watching behaviour such as the proportions actually watched and parts of videos that had more interactions, the number of publications decreases to a handful. The increased adoption of MOOC courses in universities certainly increased an interest in deeper understanding of the issues of online learning (e.g. decreased motivation, delayed feedback, feeling of isolation, etc. \citep{Yusuf2013} \citep{Coman2020}) and of the analysis of video material usage and student behaviour, the bulk of the latter will be  reviewed in this section. We expect a more recent line of such studies to appear soon as the COVID-19 restrictions in universities resulted in more intense e-learning thus an increased amount of video-based online classes.

\subsection{Within-video usage analytics in educational context}

Video usage analytics outside the educational context have been many and diverse including market surveys on why people re-watch the same commercial video content \citep{Bentley2016} and market studies using the commonly used ``views" and ``likes" counts in YouTube. The use of ``views" in particular, is the most basic, dominant metric used in learning analytics studies where the main focus is the usage of the whole learning management system (LMS) in which the video usage is a part (e.g.  \citep{Jovanovic2017}).

Other than views, there are other metrics typically used for video analytics including ``play rate" (ratio of playing a video to the number of links usually in the form of a thumbnail frame image, shown to the viewer) and ``view length" (how long the viewer watched the video). In the context of educational videos and pre-class asynchronous learning materials, ``punctuality" is sometimes used to measure whether the students watched the video before the class time. Common metrics used to look into within-video watching behaviour include ``coverage" (ratio of playback portion to the whole duration of the video), ``peak" (most frequently interacted part of the video) and ``drop-off point" (where the viewer stopped watching in the video). In the video usage analytics in the educational domain, some combinations of these metrics have been used. 

An in-depth interview with 16 post-secondary instructors who use videos in their teaching \citep{Fong2019} showed that a more in-depth analysis of video watching behaviour of students would be very useful for them to assess the video quality and how they could revise accordingly. It is apparent that lecturers will benefit if there are more detailed information available to them on how video materials are viewed by students.

On the one hand, the majority of the literature on video playback interaction analysis to support such insights in an educational context comes from around 2013 when large-scale MOOC courses became popular  and a greater  variety of in-video interaction types (pause, seek, speed change, etc.) became the norm on these platforms. This meant  that the analysis of these click-level interactions became technically feasible in practice. On the other hand, the degree of detail that the studies look at within-video interaction varies, from simple view counting presented over time to  second-by-second cumulative calculation by different interaction types.

For example, one study \citep{Dazo2016} looked at the usage of 25 video materials across 3 consecutive semesters in a flipped classroom setting. Using views, coverage and punctuality as the main metrics, they found that  students' academic performance was connected to some of the video viewing behaviour, e.g. previewing the videos before the class. Looking at the watching behaviour inside the videos was through the metric of coverage, generally proportional to the view counts. Another study found out 2 distinctive clusters of students based on their video watching behaviour in a 15-week course on Matlab programming \citep{Moore2021}, where the high-performing cluster of students also showed significantly higher coverage in their watching, compared to the low-performing cluster. 

Moving beyond the overall view count and coverage, an analysis of a MOOC course involving 48 videos  \citep{Atapattu2017} counted the number of video interactions (play, pause, speed change, etc.) that happened at each second in each of the videos, presenting the counts on a timeline within the video. While their focus was to use video transcripts to analyse the discourse and correlate with those video interaction counts on the video timeline, the counting of interactions at each second of the videos and seeing the highs and lows of them on a timeline of each video is a promising approach to reveal the minute details of interactions as they happened while the videos were playing.

Going even deeper into within-video interactivity, Li \textit{et al.} related the frequencies of different types of within-video interactions (pause, skip, replay, etc.) to the understandability of video contents (i.e. difficulty of learning) \citep{Li2015a} \citep{Li2015b}. Of particular interest is the metric ``replayed video length" that summed all played video durations by seconds that were re-watched (concluding that less re-watching indicated higher video difficulty). Differentiating different types of interactivity is certainly necessary if we want to understand the playback usage in more detail, and we describe later how in our study the different types of interactions will be assigned to different weightings in cumulative scoring inside each video.

A study in \citep{Kim2014} used similar second-by-second bin counting where play and pause interactions increased the count thus forming the ``peaks" in the usage timeline. They also used simple image similarity comparison between frames from every 1 second in the video to identify the points where high differences appear and thus  help categorise different types of transitions in video presentation (e.g. from PowerPoint slides to talking head). They show a prototype user-interface where different types of interactions (views, replay, skip, etc.) can be selected to see its cumulative peaks within the video.


Further related studies and analyses include video analytics that used aggregated time series to present  interactivity within the videos. This was used to check those parts of video contents in a specially designed 10-week course  \citep{Giannakos2015}, correlating the video playback time with student engagement level in 4 edX courses that happened in the same semester \citep{Guo2014}. This took a cognitive psychology angle where grouping of different sequences of video interactions into higher-level ``behavioural actions" to help predict the students' usage behaviour \citep{Sinha2014}.

Table~\ref{table:LiteratureTable} shows a summary of 7 video usage analysis studies (2014 - 2021) some of which addressed the above, selected as good representative studies in the literature that we reviewed. It can be observed from the table that:

\begin{itemize}
    \item Educational video contents span from a few dozen to a few hundreds of videos, and the length of each video varies from a few minutes to 45 minutes;
    \item Various sets of metrics are used in different studies;
    \item There are attempts at correlating the watching behaviour to some overall score (e.g. grades);
    \item Temporal mapping within videos (the last 3 rows in the table),  attempt at within-video locating of usage behaviour, could be explored more.
\end{itemize}

The studies that apply the last point (temporal mapping within video) count  video interactions at second-level points inside each video, including clicking of play/pause buttons, replay, etc. This allows them to show the cumulative effects of usage as the peaks of these interactions along the timeline. The peaks then point to the video contents, at those points.

It is this temporal mapping of within-video usage that our paper  builds upon, where the values to be presented are not the counting of interactions but the accumulated playback scores that take into account various video interactions, each weighted differently, in order to reflect the way students actually watched the videos.  We expect this will better guide  students as to where to watch.

\subsection{Within-video recommendation and timeline visualisation}

Typical usage of video content has been to linearly watch from start to finish, whether short  (e.g. a 5-minute YouTube clips) or long  (e.g. a 2-hour movie), thus not requiring any elaborate visualisation scheme to help to navigate or jump within the videos other than indicating where the current point of playback  is in time. Educational video content, on the other hand, will generally benefit from features which support within-video navigation/guidance as the videos in educational domains typically have a structure, a progression and different types of presentation, and students might be looking for certain parts within them. \textcolor{black}{There are cases where more use of video control features such as skip/pause was negatively associated with students' performance (e.g. \citep{Joordens2009}), though that work pre-dates the huge uptake in online learning over the past decade and in particular as a result of COVID. Thus thinking that providing various interactive video playback features will be beneficial for students may be only an assumption. Different learning strategies used by students (e.g. simply memorising vs. deeply understanding) along with the topic types of the course will have impact in deciding how beneficial the video control features are  \citep{Le2010}.}

Some of the video analysis studies in educational contexts, including some of the earlier-mentioned ones, do feature visual guidance on the playback timeline. For example, TrACE UI shows the portions of video that they already played \citep{Dazo2016} on its video timeline, guiding  students on where to play or avoid playing; Video timeline shows peaks and highlights that are dynamically changing as the usage increases \citep{Gajos2014}. YouTube's video analytics page features a ``key moments for audience retention" section in which the usage frequency is presented on a timeline (duration) of a video, visually showing the peaks and troughs presented based on  playback as well as other interactions such as sharing with others at those points. To support video-based learning of how to use Photoshop effects, the video timeline shows the highlight points for step-by-step navigation \citep{KimToolScape2014}. 

Although not particularly for an educational context, Video timelines with bookmarks  allows smooth pre-fetching and the indication of buffering has been proposed \citep{Carlier2015}; Ice hockey goal events are highlighted and shown as mini icons on the timeline of the playback UI \citep{Muller2010}. Many experimental video retrieval systems in the field of multimedia (e.g. the annual Video Browser Showdown showcase \citep{Schoeffmann2019}) have been experimenting with more sophisticated and advanced within-video highlights where the retrieval unit is sub-video parts (e.g. a frame, camera shot, scene, etc.) and the searcher is supported by various visual strategies to highlight within-video match points within long videos. Examples include highlighting  the portions of a video timeline for a particular type of scene (dialogue, action or montage) in movies \citep{Lehane2006}, indicating the boundaries between short video chunks in a longer, concatenated video \citep{Lee2017}, and various ways of highlighting retrieval results on the video timeline \citep{Gaughan2003} \citep{Cooke2004} \citep{Leibetseder2018} \citep{Sauter2020}. 
We do need  to see yet more effort in applying these sophisticated within-video timeline visualisations in an educational context. Judicious applications of some of these features could help students more efficiently learn the contents by more selectively picking and playing the parts of videos that are useful for them.

\section{System Description}

The system we built and deployed is a web-based archive of educational video materials made available throughout a semester, used in conjunction with the university-wide Learning Management System. The video materials, once they are available on the system for watching, can be played with  conventional playback-related functions. While all user interactions throughout the web interface are logged, the system's visual guidance feature which this paper  introduces and the analysis used to drive its operation, uses mainly the playback-related interaction portions  of the logs such as play, pause, and speed change as well as indirect, contextual usage logs such as whether the playback is happening as the window of focus on the student's screen. 

\begin{figure}[h!]
    \centering
    \includegraphics[width=0.9\textwidth]{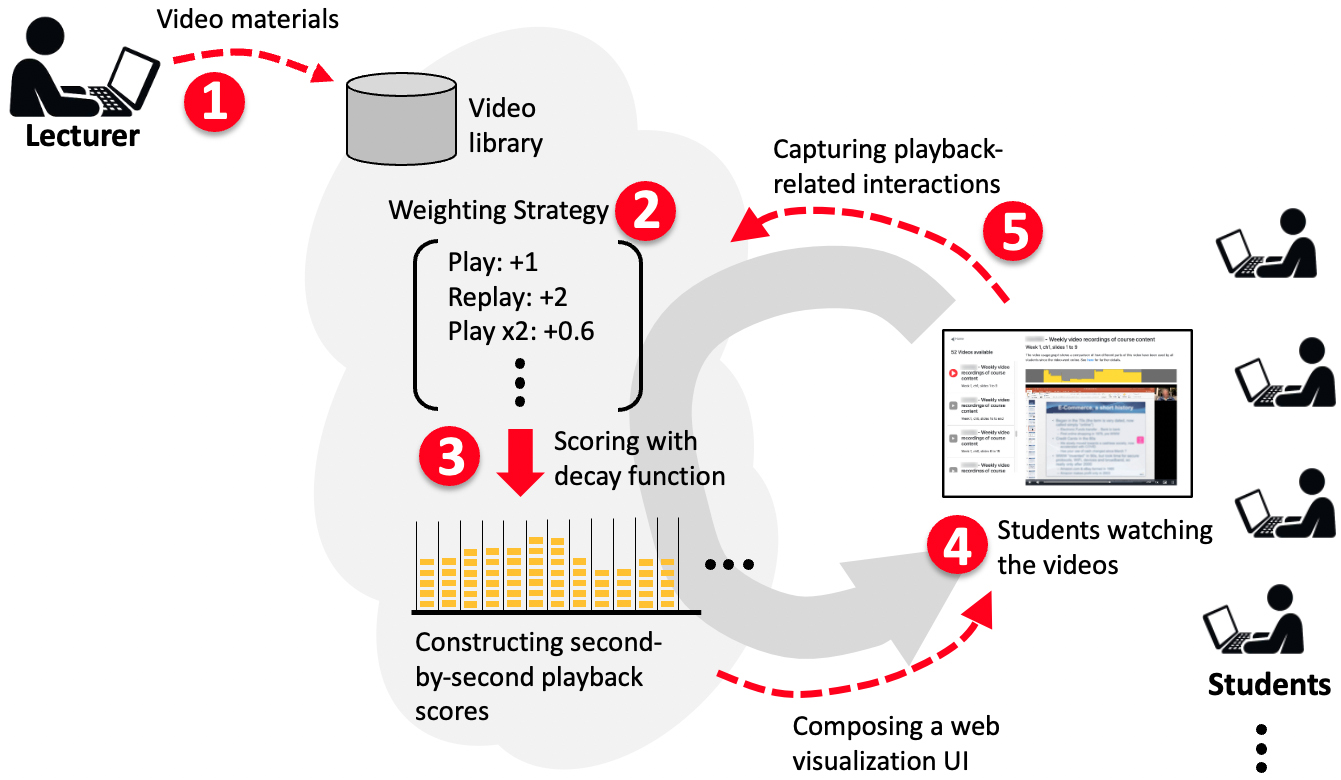}
    \caption{Diagram of how the system works: once the video materials are uploaded (1), the remaining processes in the (2)-(3)-(4)-(5) cycle iterates as the resulting visualisation contour evolves over time.}
    \label{fig:diagram}
\end{figure}

Figure~\ref{fig:diagram} shows how the system works.  The lecturer  uploads  video clips that s/he created and offers the students to watch as part of the course material (1); the system will use a specially formulated playback weighting strategy to calculate what the score for each of the second-by-second bins in the timeline should be (2); scoring adds a time-decay function to construct the playback visualisation for  students (3); the resultant UI is then presented to  students (4); finally, students  interact with the system by playing, pausing, changing speed, etc. and this interaction is fed back to the system to re-calculate the scores using the weighting strategy (5). The cycle (2)-(3)-(4)-(5) continues through the semester as the shape of the timeline visualisation  evolves over time. 


In the remainder of this section, we describe the user interaction and  visualisation, followed by the mechanism used to generate it.

\subsection{User interaction and the usage visualisation}

\begin{figure}[h!]
    \centering
    \includegraphics[width=0.8\textwidth]{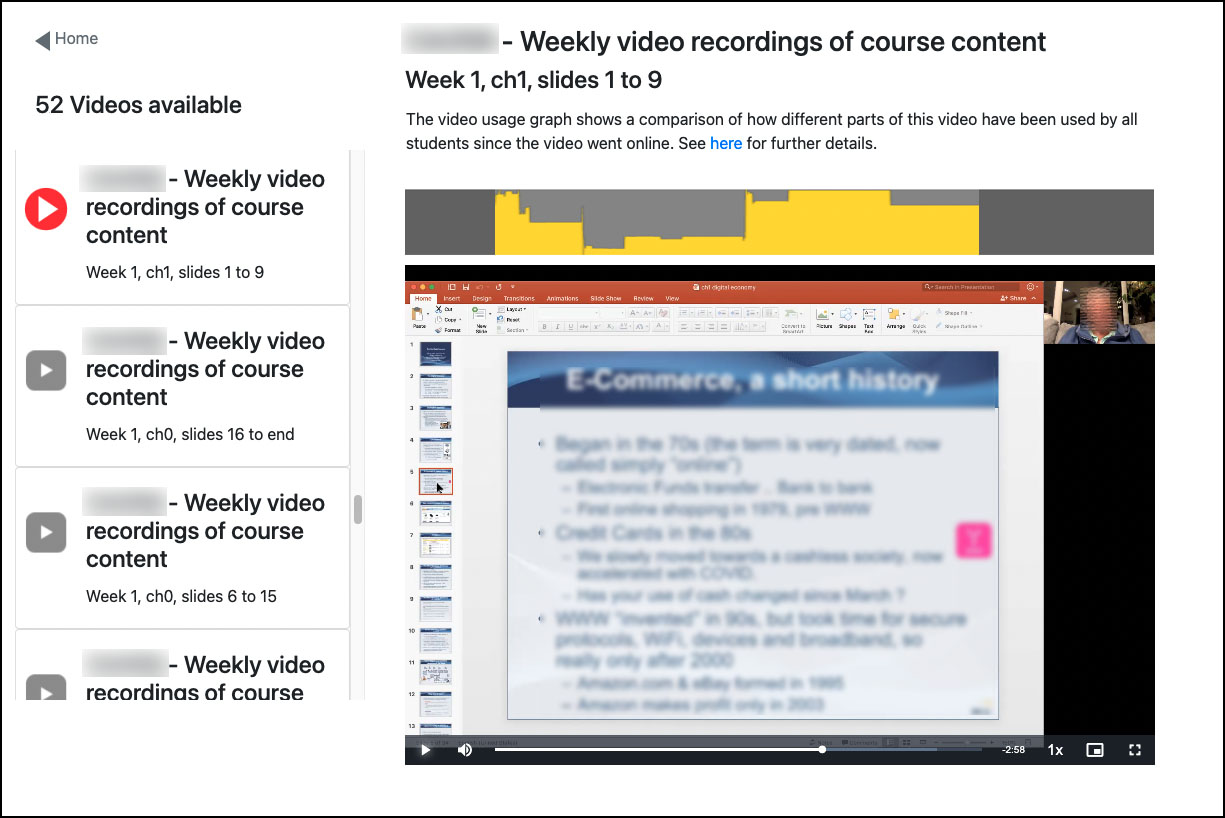}
    \caption{Screenshot of system interface showing the user has 2:58 left to play at 1x speed of what appears as a 10-minute video from week 1, chapter 1, slides 1 to 9 of the course. The yellow graph just above the video playback panel indicates the section the student is about to play has had highest usage scores based on previous video playback from the class whereas the part of the video at about the 1/5 mark has low usage.}
    \label{fig:screenshot}
\end{figure}

The web interface allows  students to log in by using their student email address and a unique 8-character code for the course. Once logged in, the student is presented with a list of available video materials on the left side of the screen (see Figure~\ref{fig:screenshot}). Selecting one  will load the video playback panel and the associated usage visualisation above it.

The visualisation feature is a time-based graph, aligned with the playback timeline of video content (panel above the playback screen in Figure~\ref{fig:screenshot}), where the height of the yellow shape suggests video usage to  students so that they could watch those portions within videos that show high levels of usage by their peers. 


Using the yellow contour of the graph as a hint to select  parts of the video to watch,  students engage in selective playback of the video, data from which will, in turn, be taken into account using the playback weighting strategy to contribute to the shaping of the contour for other students. Establishing the usefulness of this visualisation in helping  students watch the important parts of videos is the focus of the analysis  presented later.

\subsection{Playback weighting strategy for the visualisation}
\label{weighting_strategy_section}

The height of the yellow areas within the visualisation is the result of our playback weighting strategy, the core technical component of the system.

Each video is divided into 1-second windows and each window starts with an initial score of 0 which is incremented every time any student plays or skips it. Every time any part of the video is played, that part will gain a score increment, thus over time as the usage increases the scores in each second-by-second window on the timeline will increase. 

Since there are various playback-related interactions a user can perform at any time during video playback, the system uses the type of interaction that happened  to calculate the score increment for that second: thus the playback weighting strategy determines the shape of the visualisation, eventually serving as the guidance to students on what parts of the video to watch. The strategy is as follows:

\begin{itemize}
\item \textbf{Play}: As the most basic scoring strategy, the window gets \textbf{+1} when that portion is played. This is similar to calculating ``peak profile" in which play events are counted to show the peaks \citep{Kim2014}. However, if the video playback window was not the window of focus on the student's screen when the student was playing it, then this increment is \textbf{+0.25} only,  bringing down the score for  cases where the student may be reading email or doing something else while half-listening to the video.

\item \textbf{Replay (Seek backward, i.e. within the same session)}: All 1-second windows between the previous playback point and the point it skipped back to will each gain \textbf{+2} score, so each gets a cumulative +3: +1 from the initial playback and +2 from the replay). Re-watching could mean more difficulty in understanding that part of the video \citep{Li2015a}, although there could be other factors e.g. returning to  missed content, following the steps in tutorial-type videos, hearing again non-visual explanations, etc. \citep{Kim2014}. Whatever the reasons, the parts which students replayed are  parts to watch out for and  point out during students' course review. 

\item \textbf{Faster play:} Playback at $2\times$ (double speed) will gain \textbf{+0.6} and if the window is not the window of focus then it will be \textbf{+0.2}  only because when the student is attending to another window, the double-speed playback is too fast to properly comprehend. Playback at $1.5\times$ will gain \textbf{+1.5}, and \textbf{+0.5} if the playback window is not the window of focus. In this study we assume that playing at $1.25\times$ speed will gain the same as playing at $1\times$ speed for student learning. Recent studies consistently show that playing  educational videos at $1.25\times$ resulted in better learning outcomes than when played at normal speed \citep{Lang2020} \citep{Mo2022}, and the benefits and drawbacks of other speeds (e.g. $0.75\times$ and $1.75\times$) are worth further investigation in order to incorporate them into our weighting strategy.

\item \textbf{Skip (Seek forward)}: There can be different reasons from a cognitive processing point of view as to why students want to skip familiar parts of a video or continue watching even if already familiar \citep{Sinha2014}, but skipping typically implies that the student finds those portions easier to understand \citep{Li2015a}. If a student skips forward from the current position at minute S0 by a segment of video then windows within  1-, 2-, and 3-minute  segments directly following the  segment S0 will get score adjustments as follows: S60: \textbf{-0.3}, S120: \textbf{-0.2}, and S180: \textbf{-0.1}. The rationale for this deduction in scores is that the student must have had an idea what was coming up next, after point S0, but anticipated it as being not interesting or useful for her/him at that point in time and thus less likely for other students to also find it   interesting or useful.

\end{itemize}

\noindent
\textbf{Adjusting the score over time}: Overarching the playback weighting strategy is a \textbf{decay function} based on recency of playback, with the most recent days of playback interaction being more meaningful or useful than prior to that. The scores calculated by the above strategy are re-calculated from the interaction log each midnight. In this way the score increments (both + and -) as above are called our BaseIncrements = (+1.0, +0.25, +2, +0.6, +0.2, +1.5, +0.5, -0.3, -0.2, -0.1) and the system makes those the actual increments on day 0. Then on day 1 it makes those increments each multiplied by 1.1  before adding, on day 2 the base increments  multiplied by 1.2, and so on.  The effect is that on day 10 we have a score for each 1-second window which has a 10-day linear decay function so that something played on day 9 has twice the value of something played on day 0 and by day 20 we have a score which has a 20-day linear decay with the half-life being 10 days.  This  continues  indefinitely. The decay function is useful for the lecturer who wants to understand the factual usage statistics of each video segment, but it serves an important role that will influence the shape of the usage visualisation for students who need to be selective at the time of reviewing the video.

The scores for the 1-second windows are normalised within each video usage graph at display time making the visualisation less susceptible over time to any maligned attempt at artificially inflating scores by jumping to or repeatedly playing an obscure segment of video.

A study found that the points in videos with more video interactions indicate those parts requiring higher-order thinking and more cognitive skills by students \citep{Giannakos2015}. Instead of scoring highly those parts with more interactions regardless of the type of interactions, our weighting strategy attempts a more fine-grained differentiation amongst different video interactions and uses the cumulative scores from it as  overall guidance to offer to students.

\section{Course Outline, Deployment and Data Captured}
\label{outline_section}

The course for which the system was deployed was an undergraduate-level course on data analytics for marketing applications delivered in the Spring semester (18 January - 15 May) in 2021, and again  the following Spring semester (10 January - 26 April) in 2022.

Since the Spring semester in 2021 was a lockdown period due to the COVID-19 pandemic, the course was delivered fully online: Zoom lectures were conducted live while video recordings of them were made available for students for later asynchronous review as they wish. The University's virtual learning environment was also used to present other course material including links to YouTube clips, news articles and other websites as well as the PowerPoint slides used for the online lectures and for creating the video materials and worked examples of data analytics. As an additional source of  course material, the Professor prepared  a set of 52 short form video clips for students to watch in their own time in addition to various links to the other  materials. In each week, the materials suitable for that week's lecture topic became available as entries on the course's LMS site. The number of students enrolled to the course in the first deployment was 131.

The Spring semester in 2022 was conducted with regular face-to-face in-class lectures as the main mode of course delivery and those lectures were also streamed live and Zoom recordings were made available, further augmented with the same additional online material as in the previous semester including the same set of video clips prepared by the Professor from 2021. Each of the video clips was pointed out and recommended for students to watch as the course progressed through the face-to-face lectures. The number of students enrolled in the second deployment was 139.

The video materials were screencasts created by the Professor who taught the course. Each video was about 10 minutes duration and 3-5 such videos were added for each week throughout the semester. \textcolor{black}{The topic for each of the videos is shown in Table~\ref{table:Topics}.}  The time of the addition of each video depended on  factors including the availability of certain materials to be mentioned in the video and time constraints of the Professor, resulting in some materials on later topics to be available earlier in the semester, and vice versa. A total of 52 videos had been made available to students in both deployments, and the usage of these 52 videos facilitated by the visualisation during the two deployments is our main focus in this paper.

For the duration of each of the two semesters,  enrolled students used the system actively playing and re-playing  video content as part of their studies thus feeding into the playback usage analysis which, in turn, helped shape the contours on the timelines for each video that guided them where to watch.

A manual annotation of the most important part within each of the videos was carried out in order to check whether the visualisations actually helped students locate the most important parts of videos in their study. The Professor who developed the video materials and delivered the course went through each video, marking the start and end points within each video that were judged as the most important part of that video, not knowing the actual playback usage that happened during the semester. The only constraint imposed on the Professor's annotation was that for each video there was to be only one continuous part as the most important, that is, not multiple different time blocks. The duration of this part was up to the Professor to decide purely based on the learning content. Although there might be multiple points that a lecturer sees as important in courses where video materials are longer in duration (e.g. 30 minutes or 1 hour or more), the videos in this course were short and the Professor specifically designed each video to be about one particular concept. Since the course and its video materials were all developed by the same Professor, there was no need to employ multiple assessors to indicate these important parts and to then check cross-annotation for consistency.

It is worth noting that the annotation of the important parts within each video is inevitably a manual task that had to be performed by the Professor in order to answer the research question in our study, and thus for the regular deployment of the system such a manual effort is not required.

\section{Findings and analysis}

Overall usage of the video materials using our system is summarised in Table~\ref{table:overallusage}, showing  combined usage from both semesters. The analysis and our evaluation of the usefulness of the proposed playback weighting strategy and visualisation uses the combined usage data collected from both semesters. Since the video clips were supplementary to the main lectures by the Professor either by Zoom or face to face or both, as well as to other online materials,  students were not required to watch all available videos (though they were encouraged to), which explains the overall low playback usage (an average 28.9 minutes of playback per student). This indicates that when students did use these videos it was not their main source of course material and could have been used to clarify some complex aspect of the course, though this is speculative.

\subsection{Playback usage according to playback weighting}
\label{playback_usage_subsection}

\begin{figure}[h!]
    \centering
    \includegraphics[width=1.0\textwidth]{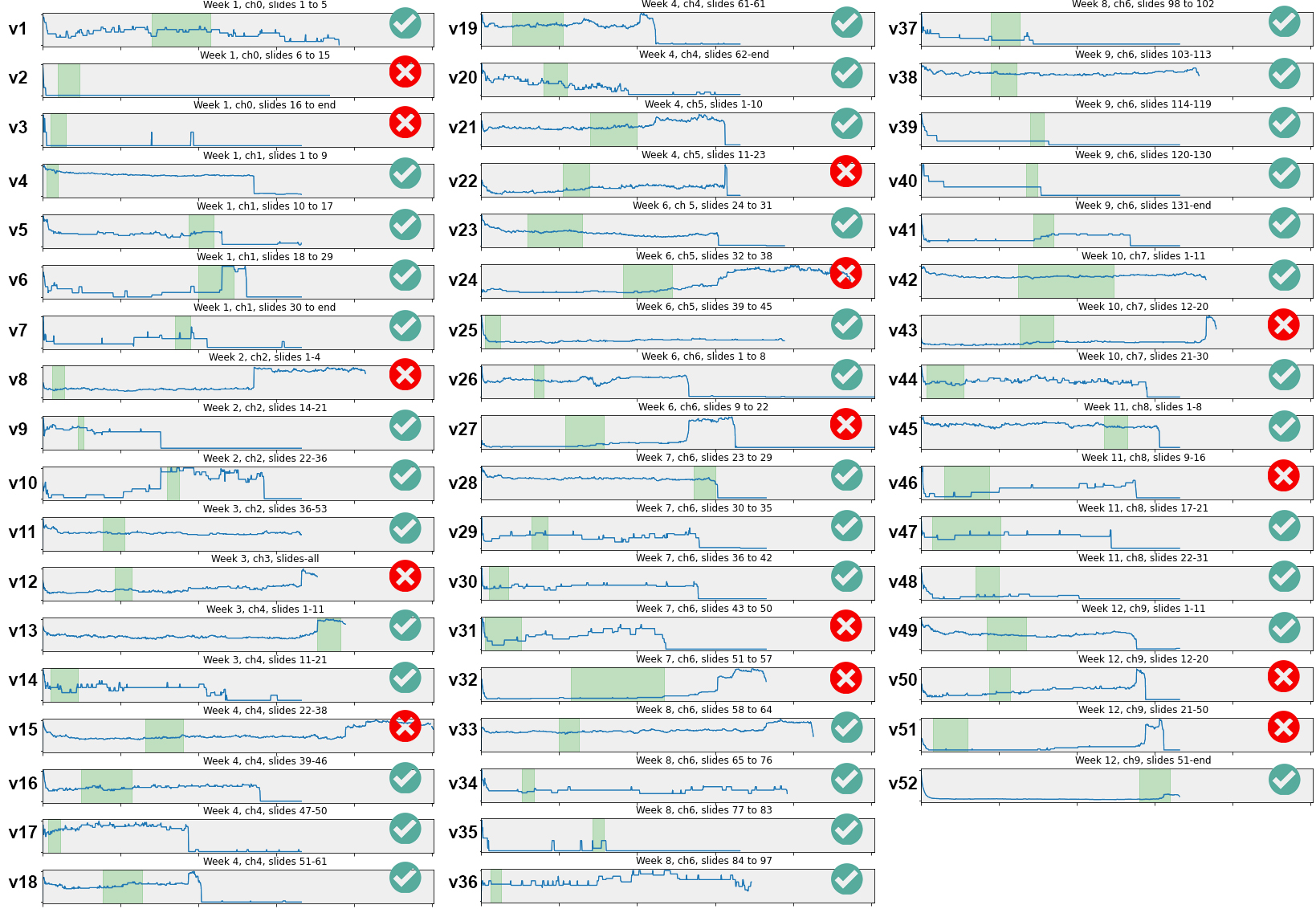}
    \caption{Within-video playback usage for each of the 52 videos  using the playback weighting strategy (without decay function), each video labelled v1 - v52. In each graph, x-axis is video duration, y-axis is a normalised overall playback score for each second. In each video, highlighted in light green is the most important part in the video; \checkmark  or $\times$ symbols indicate whether the highest plateau covers the most important part (i.e. students watched the important part) or not.}
    \label{fig:Minutes}
\end{figure}

In order to answer our research question, we examine the playback usage according to the timeline visualisation shaped by our playback weighting strategy. Figure~\ref{fig:Minutes} shows the playback behaviour within each of the 52 videos according to our weighting strategy but without reflecting the decay function (designed to factor in the date  at which the student is watching the video).
%
%
Each graph represents the second-by-second playback scores presented for each video, according to the playback weighting strategy  described in Section~\ref{weighting_strategy_section}.

The blue contour lines in each video graph show the playback usage of each video. As can be seen, the contours are overall quite different-looking than those that present the counts of video interactions since here we present the cumulative, played parts of videos with the influence of the playback weighting strategy: there are fewer sharp peaks but a greater number of different levels of plateaus where  playback continued for some time.

To begin with, an initial observation from Figure~\ref{fig:Minutes} is that all graphs start with  high playback usage. Students start watching a video by clicking on the play button, and as the starting point is always at the start of the video when the video is first loaded, the playback score is always counted at the very start of every video. But after  playback starts, almost always the students will skip into the middle of the videos, making the playback scores at the start immediately drop. Exceptions to this behaviour can be found in only  a few videos, for example, the video labelled v45 in Figure~\ref{fig:Minutes} shows that from the very start,  playback usage was well maintained for  most of the video; the video labelled v31 shows a start with a low usage but steadily increasing until the end of the video at around the 5 minute point.

The timeline visualisation that students actually saw and were guided in choosing the portions to watch are the decay function-enhanced versions of the Figure~\ref{fig:Minutes}, that showed slightly different contours depending on when they watched the videos. It is not possible to accurately estimate how the usage might have been without the decay function influencing the visualisation as the usage evolved over time with the function, but by back-tracing the contours during the semester, we can estimate its effect (see Section~\ref{change_section}).

\textbf{Effectiveness of the visualisation: highlighting the important parts.}
\label{effectiveness_section}
Checking the playback usage together with the important parts manually annotated by the Professor is a central task to answer our main research question. In Figure~\ref{fig:Minutes} the green areas in each video graph are the important part in each video marked by the Professor who gave the course and annotated this independent of the actual usage. 
The marking range for each video was then compared to the playback usage visualisation for each video, to check whether the parts with high plateau of playback scores (i.e. the part the students were guided to and watched) is where the Professor marked as the most important parts. The green check symbol on the right side of each video graph in Figure~\ref{fig:Minutes} shows that the highest playback regions matched the important parts of the videos, while the red 'X' symbol shows that they did not match. Out of 52 videos, 38  matched (73.1\%): that is, 38 out of 52 playback visualisations correctly guided  students to watch the most important parts within the videos and they actually watched those portions.

This deserves  more detailed analysis. For example, video v6 (see Figure~\ref{fig:Minutes}) shows a low usage for the first 5 minutes or so, then suddenly peaks maintaining the plateau for 30 seconds before flattening, and the peak point coincides with the important part as judged by the Professor. It is encouraging to see that the natural playback usage reinforced by the visualisation, results in  students watching that important part of the video. As another example, v41 shows relatively low usage throughout, but halfway into the video at around 3 minute point, the line slightly goes up and remains plateaued: the start of that plateau is the important part and students correctly catch that part.

On the other hand, v42 shows a middle third of the duration as the important part, yet the actual playback was pretty much a high plateau all throughout the duration of the video. The visualisation ends up not differentiating any part and is thus unable to effectively guide  students to be more selective, yet most students who opened this video did watch the important part.

\subsection{Contour changes over time}
\label{change_section}

\begin{figure}[h!]
    \centering
    \includegraphics[width=0.7\textwidth]{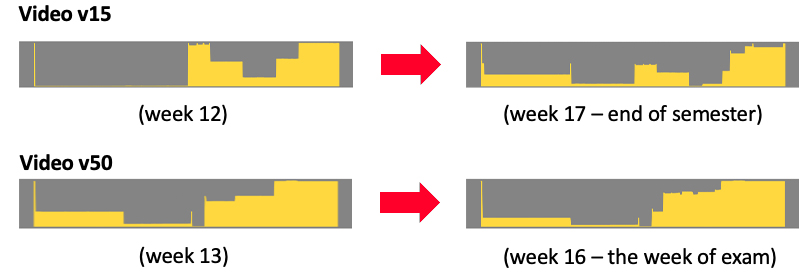}
    \caption{Timeline changes over time. v15 (duration: 19m 02s) at week 12 and just after the final exam; v50 (duration: 9m 34s) at the week of final exam.}
    \label{fig:timeline-change}
\end{figure}

As  students play the videos, the timeline visualisation for each video changes  as it is updated daily. Figure~\ref{fig:timeline-change} shows two example timelines  captured near the end of lecture weeks (Weeks 12 and 13 respectively) and again at/after the final exam, during the 1st deployment (since the visualisations that the students actually saw cannot be based on aggregated data across two semesters).  In case of v15, \textcolor{black}{the topic covers the normal distribution in a set of numeric values and the concept of how the standard captures the spread of data values.  The parts  played most are the the second half by the end of lecture classes as students grapple with the concept of what is a standard deviation.}   Approaching the final exam the playback portions changed and the week after the exam the timeline shows more coverage across the video, ending up with two moderate plateaus at the start and middle followed by the highest region nearer to the end of the video. 

On the other hand, v50 \textcolor{black}{which is about how our digital footprints can be used to predict our personality using the 5-factor personality model}, this seems to maintain the overall playback usage after the high use at the exam, though slightly lower at the first 3 minutes or so, and higher at the latter 4 minutes. 

\textcolor{black}{The change of contours becomes more noticeable as the usage becomes higher, and in our system, this understandably occurred right before the final exam period started. Whatever contours shaped thus far as a result of relatively low but continuous and gradual usage during the main part of the course  are then picked up as the hints to where to watch when the urgency of learning increases at the impending final exam (the danger of ``bandwagon effect'' and how our algorithm is resistant to it is discussed in the next section). However, the exact pattern in this more noticeable contour change near the exam time is not consistent and dependent on the video contents, as seen, for example, the way the peaks near the end becoming more spread out throughout v15 as the result of more usage due to final exam, and maintaining the peak near the end of video and becoming increasingly so due to final exam in v50, clearly contrasting to v15.}

\section{Discussion}
\label{discussion_section}

\textbf{Playback-centric view as complementary to video interaction view.}
The contours of the within-video graphs are quite different from those that present the counts of video interactions (e.g. \citep{Kim2014} \citep{Giannakos2015} and \citep{Atapattu2017}). Influenced by the playback weighting strategy that incorporates  video interactions but presented not by those interaction points but by the played parts, the playback-centric presentation is more range-based (i.e. from time A to time B) than point-based (i.e. this point in time). The message the visualisation  gives is the suggestion of the parts of the video to watch and is thus slightly different from the presentation of where the interactions occurred most often. The playback duration-based usage on the timeline is an additional aspect of the usage which is complementary to the interaction point-based usage, adding further dimensions to  understanding  the details of the use of video playback.

\textbf{High plateau = important part?}
In this study, we attempted to assess the usefulness of the visualisation (and its weighting strategy) with the broad assumption that the system's visualisation is effective when the most-watched parts in videos correspond to the most important parts as determined by the Professor. This assumption may be naive since the parts where the students need to review more frequently or need to think more deeply might not be same as the most important parts identified by the Professor.  Perhaps the more difficult parts, or the parts that omitted examples or were explained poorly, might have been watched more and thus the more difficult or poorly-explained parts are not necessarily the most important parts. However, as an initial deployment of the visualisation, such an assumption could still be useful without complicating or adding further intervention steps during and after the deployment. This needs to be kept in mind in the interpretation of our findings.

Currently our playback weighting strategy is purely based on  playback usage and influenced by the time-decay function. For 38 out of 52 videos in our study, the parts the students watched the most were largely the parts that the Professor judged as most important, but for the remaining videos this was not the case. The weighting strategy could incorporate content-related factors e.g. the instructor's indication of important parts in each video, perhaps by ranking a few different parts of the videos, to influence the baseline score of each second-by-second bins. However, the added manual input required from the course instructor would be an issue especially as the number of courses and videos increases. 

The timeline visualisation changes over time as  students continue to play different parts of different videos. Students will not benefit from the visualisation in the early stages of the semester as not many parts will have been played and not a lot of noticeable contours will be shown. They benefit as the timeline plateaus become more prominent, only after it becomes clear where peaks are and this   comes only after  playback usage has occurred. 

With this in mind, we now ask whether the proposed visualisation helps 
students selectively watch more important parts within the videos and the answer is reasonably so (73.1\%), considering no explicit information or intervention was offered to influence video usage. Furthermore, given our approach of relying on usage data to feed into the system, the visualisation is likely to improve its guidance to important parts if there was a pedagogical mechanism to increase the overall usage by students e.g. enforcing the watching of the videos as mandatory, rather than as supplementary material as  in this course. 

What happens when the high plateaus are not corresponding to important parts of the video? If so, it could mis-guide the students and potentially fall into a negative bandwagon effect as happens in social networks \citep{Wang2015}), in recommender systems \citep{Sundar2008} and perhaps take advantage of the effect in e-commerce and IoT products \citep{Choi2015}. Unintended mis-guiding formation of the usage visualisation is less likely to occur as the decay function embedded in the playback weighting strategy makes the shaping of the contours somewhat resistant to occasional random watching on obscure points within the videos. Initial ``seeding" of important parts in each video may be a way to further reduce such negative effects.

\textcolor{black}{Finally, the videos watched by the students in this study were the supplementary learning material in addition to the formal online/face-to-face lectures as well as other website links and YouTube resources that were made available for students. The playback usage pattern within the videos (thus the shapes of the contours) is expected to be different if the videos are to become the core delivery mechanism as is the case in many studies referenced in this paper so far (e.g. majority of the usage analytics studies and the ones in Table~\ref{table:LiteratureTable} are MOOC courses in which the video playback is the central vehicle for content delivery). The interpretation and the application of our findings and discussions on the behaviour of the visualisation and its effects for other courses, therefore, will need to take this into account.}

\textbf{Future studies.} This study was the first deployment using the strategy that weighed individual interactions differently. One reason for pushing for  full deployments integrated into an existing course was to identify which aspects of the system could or should be the main focus of  future studies within the design, implementation and deployment, and also how the system and the way the course is delivered could more tightly support each other.

There is  room for improvement by incorporating other more detailed playback contexts into the playback weighting. For example, the faster/slower speed setting could have more elaborate weighting variations depending on the target speed set, informed by other studies on the particular speed affecting the students' overall performance; other interaction factors that could be used include the   volume level used while playing and whether it was viewed full-screen.

While in this paper we focused on understanding the overall usefulness of the visualisation and its potential value, more detailed analysis specifically delving into different categories of session behaviours or student group behaviours with statistical means will allow us to characterise its usefulness in more specific ways. Similarly, designing a comparative study which has a half of the student user population with the visualisation and the other half without it or with a different weighting strategy will help understand the level of usefulness of the visualisation and its weighting parameters.

One of the potential issues inherent in our approach is when each video is first uploaded and  becomes available on the system without any prior playback usage data. The ``cold-start" issue is typical of many usage-based machine learning algorithms such as recommender systems (\citep{Natarajan2020} \citep{Kang2019}) where the effectiveness of the approach relies on its usage over time. In our system, when a video first becomes available, the visualisation does not have  usage data from which to calculate and populate  the timeline. As soon as a student starts using a video, its usage will be captured and reflected as a low, yellow area at the bottom of the timeline. A manual indication of the important parts by the lecturer at the time of uploading each video may be a simple solution that could also serve as a way to lead a more desirable evolution of the contour. However given one of the strengths of our approach is  fully-automated shaping of the visualisation without  extra  efforts by a human, other more automated ways to draw the data will be an important topic of further study. For example, some studies used supervised machine learning techniques applied to lecture video images to determine the important parts in  videos \citep{Brooks2009} or applied to the usage data (selecting a video, seeking within the video, as well as the heartbeat rates of the students) to develop a computational model of the usage patterns of students to predict a more desirable pattern \citep{Brooks2014}. Taking a bottom-up indexing approach (i.e. important parts emerge by looking at the video/usage data) rather than top-down (i.e. system first defines and codes in what the important events are), bringing some of these computational approaches into our system may help tackle this issue in a way aligned to what we eventually envision. In our second deployment, the same videos used in the first deployment were uploaded afresh without any use of the data captured from the first deployment. One possible approach would be to plug in the usage data from a previous running of the course (if it has run in  previous year(s)) at the start of the course, to be slowly overridden by the actual usage once this was sufficient.

More refined versions with a tailored playback weighting strategy could well be applied outside the educational domain. For example, such an indication could be useful cues in general video platforms such as YouTube and Vimeo in which the user communities already drive the access and usage.

\section{Conclusion}

We introduced visualisation guidance to students playing educational videos based on calculating a set of playback weightings for different video interactions from students occurring in different playback contexts.  The result shows that for 73.1\% of cases, those parts guided by the visualisation and thus actually watched by students, were the parts that the instructor judged as the most important parts of the videos. By placing the results of video playback usage not to the post-usage summative stage but to the middle of usage where it will help the users use the system, we  broadened the boundaries of video usage analytics, starting with an educational setting where more explicit within-video guidance is very much needed.

\section*{Acknowledgements}
This research was partly supported by Science Foundation Ireland under Grant Number SFI/12/RC/2289 P2, cofunded by the European Regional Development Fund. The
research was also supported by the Google Cloud COVID-19 Credits Program for our cloud based development work.

\section*{Data Availability Statement}
The datasets generated for this study can be found in the Figshare repository  (DOI: 10.6084/m9.figshare.20288763.v2).

\bibliographystyle{frontiersinSCNS_ENG_HUMS} 
\bibliography{test}

\vfill
\pagebreak

\begin{table}
\sffamily
\tiny
\centering
\caption{Comparison table amongst 7 representative analysis studies of educational video usage (this table shows only those metrics used and applied as the main focus of the investigation in each study)}
\resizebox{\linewidth}{!}{%
\begin{tabular}{|>{\hspace{0pt}}m{0.2\linewidth}|>{\centering\hspace{0pt}}m{0.1\linewidth}|>{\centering\hspace{0pt}}m{0.1\linewidth}|>{\centering\hspace{0pt}}m{0.1\linewidth}|>{\centering\hspace{0pt}}m{0.1\linewidth}|>{\centering\hspace{0pt}}m{0.1\linewidth}|>{\centering\hspace{0pt}}m{0.1\linewidth}|>{\centering\arraybackslash\hspace{0pt}}m{0.1\linewidth}|} 
\hline
 & \multicolumn{1}{>{\hspace{0pt}}m{0.1\linewidth}|}{\textbf{\citep{Sinha2014}}} & \multicolumn{1}{>{\hspace{0pt}}m{0.1\linewidth}|}{\textbf{\citep{Kim2014}}} & \multicolumn{1}{>{\hspace{0pt}}m{0.1\linewidth}|}{\textbf{\citep{Giannakos2015}}} & \multicolumn{1}{>{\hspace{0pt}}m{0.1\linewidth}|}{\textbf{\citep{Li2015a}}} & \multicolumn{1}{>{\hspace{0pt}}m{0.1\linewidth}|}{\textbf{\citep{Dazo2016}}} & \multicolumn{1}{>{\hspace{0pt}}m{0.1\linewidth}|}{\textbf{\citep{Atapattu2017}}} & \multicolumn{1}{>{\hspace{0pt}}m{0.1\linewidth}|}{\textbf{\citep{Moore2021}}} \\ 
\hline
Course~ ~ ~ & \multicolumn{1}{>{\hspace{0pt}}m{0.1\linewidth}|}{Coursera (prgm.)} & \multicolumn{1}{>{\hspace{0pt}}m{0.1\linewidth}|}{4 edX courses (stat., AI,  chem., prgm. in 1 semester} & \multicolumn{1}{>{\hspace{0pt}}m{0.1\linewidth}|}{Specially designed course (7 weeks)} & \multicolumn{1}{>{\hspace{0pt}}m{0.1\linewidth}|}{2 Coursera courses (prgm. and sig. proc.)} & \multicolumn{1}{>{\hspace{0pt}}m{0.1\linewidth}|}{1 course (prgm.) revised for 3 consecutive semesters (15 weeks each)} & \multicolumn{1}{>{\hspace{0pt}}m{0.1\linewidth}|}{AdelaideX (edX) over 6 weeks} & \multicolumn{1}{>{\hspace{0pt}}m{0.1\linewidth}|}{Prgm. course for 2 semesters (15 weeks each)} \\ 
\hline
Class size & \multicolumn{1}{>{\hspace{0pt}}m{0.09\linewidth}|}{22k students} & \multicolumn{1}{>{\hspace{0pt}}m{0.1\linewidth}|}{128k students} & \multicolumn{1}{>{\hspace{0pt}}m{0.1\linewidth}|}{11 students} & \multicolumn{1}{>{\hspace{0pt}}m{0.1\linewidth}|}{32k students} & \multicolumn{1}{>{\hspace{0pt}}m{0.1\linewidth}|}{23,
19, 24 students each semester} & \multicolumn{1}{>{\hspace{0pt}}m{0.1\linewidth}|}{26k registered students} & \multicolumn{1}{>{\hspace{0pt}}m{0.1\linewidth}|}{215, 275 students each semester} \\ 
\hline

\textcolor{black}{Course topic} & \multicolumn{1}{>{\hspace{0pt}}m{0.1\linewidth}|}{\textcolor{black}{Functional programming using Scala}} & \multicolumn{1}{>{\hspace{0pt}}m{0.1\linewidth}|}{\textcolor{black}{Intro. CS, programming, statistics, AI \& Solid State Chemistry}} & \multicolumn{1}{>{\hspace{0pt}}m{0.1\linewidth}|}{\textcolor{black}{Use of information for problem-solving, research \& decision-making}} & \multicolumn{1}{>{\hspace{0pt}}m{0.1\linewidth}|}{\textcolor{black}{Reactive programming \& digital signal processing}} & \multicolumn{1}{>{\hspace{0pt}}m{0.1\linewidth}|}{\textcolor{black}{Intro. Java programming concepts}} & \multicolumn{1}{>{\hspace{0pt}}m{0.1\linewidth}|}{\textcolor{black}{Intro.  Processing-JS programming concepts}} & \multicolumn{1}{>{\hspace{0pt}}m{0.1\linewidth}|}{\textcolor{black}{Intro.   MATLAB programming}} \\ 
\hline

Archive size & \multicolumn{1}{>{\hspace{0pt}}m{0.1\linewidth}|}{48 video lectures~ ~ ~} & \multicolumn{1}{>{\hspace{0pt}}m{0.1\linewidth}|}{862 videos (average 7 min)} & \multicolumn{1}{>{\hspace{0pt}}m{0.1\linewidth}|}{7 videos, for pre-class preparation only} & \multicolumn{1}{>{\hspace{0pt}}m{0.1\linewidth}|}{94 videos, total 300k sessions (visits)} & \multicolumn{1}{>{\hspace{0pt}}m{0.1\linewidth}|}{25 videos for pre-class watching, avg 45 min} & \multicolumn{1}{>{\hspace{0pt}}m{0.1\linewidth}|}{48 videos, average 3.63 min} & \multicolumn{1}{>{\hspace{0pt}}m{0.1\linewidth}|}{70 videos, each 5-10 min} \\ 

\hline
Video interaction used (metrics) &  &  &  &  &  &  &  \\ 
\hline
-~ View &  & \checkmark & \checkmark &  & \checkmark &  & \checkmark \\ 
\hline
-~ Coverage &  & \checkmark &  &  & \checkmark &  & \checkmark \\ 
\hline
-~ Punctuality &  &  &  &  & \checkmark &  & \checkmark \\ 
\hline
-~ Counting
all clicks &  &  &  &  &  & \checkmark &  \\ 
\hline
-~ Play/pause & \checkmark & \checkmark &  & \checkmark &  & \checkmark &  \\ 
\hline
-~ Skip/Replay & \checkmark & \checkmark & \checkmark & \checkmark &  & \checkmark &  \\ 
\hline
-~ Speed
change & \checkmark &  &  & \checkmark &  & \checkmark &  \\ 
\hline
-~ Exit/quit &  & \checkmark &  &  &  &  &  \\ 
\hline
-~ Caption
on/off &  &  &  &  &  & \checkmark &  \\ 
\hline
-~ Play
duration & \checkmark &  &  &  &  &  &  \\ 
\hline
Differentiating among interaction & \checkmark & \checkmark &  & \checkmark &  &  &  \\ 
\hline
Clustering students/behaviour & \checkmark &  &  &  &  &  & \checkmark \\ 
\hline
Correlating the interaction to a grade/outcome & Dropout &  & Video assessment scores & Perceived difficulty & Course grades &  & Test scores each week \\ 
\hline
Temporal mapping within video & \multicolumn{1}{>{\hspace{0pt}}m{0.1\linewidth}|}{} & \multicolumn{1}{>{\hspace{0pt}}m{0.1\linewidth}|}{} & \multicolumn{1}{>{\hspace{0pt}}m{0.1\linewidth}|}{} & \multicolumn{1}{>{\hspace{0pt}}m{0.1\linewidth}|}{} & \multicolumn{1}{>{\hspace{0pt}}m{0.1\linewidth}|}{} & \multicolumn{1}{>{\hspace{0pt}}m{0.1\linewidth}|}{} & \multicolumn{1}{>{\hspace{0pt}}m{0.1\linewidth}|}{} \\ 
\hline
-~ Interaction
counting &  & \checkmark & \checkmark &  &  & \checkmark &  \\ 
\hline
-~ Visual
shifts in video &  & \checkmark &  &  &  &  &  \\ 
\hline
-~ Discourse~ ~ ~ &  &  &  &  &  & \checkmark &  \\
\hline
\end{tabular}
}
\label{table:LiteratureTable}
\end{table}

\begin{table}
\centering
\sffamily
\tiny
\begin{threeparttable}
\textcolor{black}{
\caption{Topic of each video}
{\renewcommand{\arraystretch}{1.6}%
\begin{tabular}{ll|ll} 
\toprule
v1 & Course overview & 
v27 & Trusting data visualisations \\ 
v2 & Course content, exams, assignments & 
v28 &  Data visualisations reveal data\\ 
v3 &  Online resources, past student feedback & 
v29 & Sample classic data visualisations\\ 
v4 &  Our digital economy & 
v30 & Why data visualisation is important\\ 
v5 &  ECommerce, how and why it works & 
v31 & Perceive, interpret, comprehend. Pareidolia \\ 
v6 &  Traditional marketing & 
v32 & Worked examples of human visual processing\\ 
v7 &  Digital marketing \& analytics \& ``free'' & 
v33 & Chart types, bar charts, connected dots\\ 
v8 &  What are our digital footprints & 
v34 & Radar, polar, range, box and whisper, histograms\\ 
v9 &  GDPR introduction & 
v35 & Treemaps, sunburst, bubble, sankey, line charts\\ 
v10 &  Tracking web browsing activities & 
v36 & Choosing charts to use, single numbers\\ 
v11 &  GDPR Principles &
v37 & Correlations, regression, pie charts\\ 
v12 &  GDPR details \& how people differ & 
v38 & Overview of colour, RGB \& HSL\\
v13 &  Digital footprints \& personality profiling& 
v39 & Nominal data, ordinal data \& colour\\ 
v14 &  Descriptive statistics, probability distributions& 
v40 & Data visualisation and colour\\ 
v15 &  Normal distribution, standard deviations& 
v41 & Animation in data visualisation, Tableau\\ 
v16 &  Poisson distributions, scatter plots, correlations& 
v42 & Logistic regression\\ 
v17 &  Sampling bias, significance tests& 
v43 & Machine learning overview\\ 
v18 &  Chi test, introduction to regression& 
v44 & Performance of machine learning  models \\ 
v19 &  Linear/logistic regression, regression principles & 
v45 & Market segmentation, Google and Facebook\\ 
v20 & Sample regression  & 
v46 & Algorithms for market segmentation\\ 
v21 &  Linear regression in Excel& 
v47 & Using transaction, interaction \& external data\\ 
v22 &  Data in normal form, data imputation& 
v48 & Where to find external open data\\ 
v23 &  Joining or merging data, VLOOKUP in Excel& 
v49 & Data analytics in politics\\ 
v24 &  Data Analytics pipeline / visualisation overview& 
v50 & Digital footprints to infer personality\\ 
v25 &  Storytelling in data analytics& 
v51 & Facebook, Cambridge Analytica \& US election/Brexit\\ 
v26 & Two data analytics case studies & 
v52 & Living trace-free on the internet\\ 
\bottomrule
\end{tabular}
}
\label{table:Topics}
}

\begin{tablenotes}
    \item 
\end{tablenotes}
\end{threeparttable}
\end{table}

\begin{table}
\centering
\sffamily
\tiny
\begin{threeparttable}

\caption{Overall playback usage during the two deployments}
{\renewcommand{\arraystretch}{1.6}%
\begin{tabular}{ll} 
\toprule
Number of students & 270 \\ 
Total number of videos  & 52 (average duration: 10m 08s) \\ 
Total number of sessions* & 1,521  \\ 
Total hours of playback & 130 \\ 
Average minutes of playback per student & 28.9 \\ 
\bottomrule
\end{tabular}
}
\label{table:overallusage}
\begin{tablenotes}
    \item *A session is defined as a student watching at least one video without being inactive for more than 10 minutes.
\end{tablenotes}
\end{threeparttable}
\end{table}

\end{document}